\documentclass[final,1p,times]{elsarticle} 

\usepackage{graphicx}
\usepackage{amssymb} 
\usepackage{amsthm} 
\usepackage{lineno}

\journal{Nuclear Physics A} 

\newcommand \bea{\begin{eqnarray} }
\newcommand \eea{\end{eqnarray} } 
\newcommand \nn {\nonumber}

\begin{document}

\begin{frontmatter} 

\title{Estimating $\hat{q}$ in Quenched Lattice SU(2) Gauge Theory }

\author{Abhijit Majumder}
\address{Department of Physics and Astronomy, Wayne State University, Detroit, Michigan, 48201, USA.}

\begin{abstract} 
The propagation of a virtual quark in a thermal medium is considered. 
The non-perturbative jet transport coefficient $\hat{q}$ is estimated in quark less SU(2) lattice gauge theory. 
The light like correlator which defines $\hat{q}$, defined in the regime where the jet has small virtuality compared to 
its energy, is analytically related to a series of local operators in the deep Euclidean region, where the jet's virtuality is 
of the same order as its energy. It is demonstrated that in this region, for temperatures in the range of $T$=$400$-$600$~MeV, 
and for jet energies above $20$~GeV, the leading term in the series is dominant over the next-to-leading term and thus 
yields an estimate of the value of $\hat{q}$. In these proceedings we discuss the details of the numerical calculation.
\end{abstract} 

\end{frontmatter} 


\section{Introduction}

In several formulations of jet modification~\cite{Baier:1996sk}, transport coefficients play a crucial role. 
These, in principle, non perturbative quantities encode the effect of the medium on the modification encountered by the hard jet. 
In the Higher-Twist (HT) approach~\cite{Wang:2001ifa}, 
transport coefficients represent the expectation values of operator products that have been factorized from the hard process of jet scattering and 
radiative emission in the medium. In the standard HT formalism, these are considered entirely non-perturbative and no attempt is made to 
evaluate them. In these proceedings, we outline the results of a reformulation of the HT scheme allowing for the evaluation of these coefficients 
in lattice gauge theory at finite temperature. The modification of the formalism has been described in Refs.~\cite{Majumder:2012sh}. 
Here we discuss the pertinent numerical results. 

\section{Formulation in Minkowski Space-Time}

Consider a single quark with a large negative light-cone momentum $q^{-}$ propagating through a box of length $L$, 
held at a fixed temperature $T=1/\beta$, undergoing a single scattering which imparts transverse momentum to the quark. 
The quark then exits the medium and is observed. The length of the medium will be considered to be short enough such that the 
quark does not radiate during traversal. The coupling of the jet with the medium will be considered to be small enough 
such that secondary scattering is minimal.
Dividing the square of the transverse momentum gained by $L$, we obtain the non-perturbative definition of the transport 
coefficient $\hat{q}$, i.e., 
\bea
\hat{q} = \frac{4 \pi^2 \alpha_s}{ N_c } \int \frac{dy^- d^{2} y_{\perp}}{(2 \pi)^{3}} d^{2} k_{\perp} 
e^{ -i \frac{k_{\perp}^{2}}{2q^{-} }  y^{-} +  i\vec{k}_{\perp} \cdot \vec{y_{\perp} } } 
\langle n |\frac{e^{-\beta E_n}}{Z} {F^{+,}}_\perp (y^-, y_{\perp}) F_\perp^+ (0)  | n \rangle.
\eea
In the equation above, $y^{-}$ is a light-cone separation, $F^{\mu \nu}= t^{a} {F^{a}}^{\mu \nu}$ is the gauge field strength, $|n\rangle$ is a state of the thermal ensemble with a partition function $Z$, $\alpha_{S}$ is the strong coupling constant, and $N_{c}$ is the dimensionality of the fundamental representation.

Compared to the large energy of the hard parton, $q^{-} \sim Q$ (where $Q$ denotes the hardest scale in the problem), the transverse momentum imparted 
to the parton $(k_{\perp})$ is considered to be rather small, $k_{\perp} \sim \lambda Q$, where $\lambda \ll 1$ is a small dimensionless constant. 
As a result, the $(+)$-component of the the momentum transferred, $k^{+} \sim k_{\perp}^{2}/(2q^{-}) $ is of the order of $\lambda^{2}Q$. 
As argued in Ref.~\cite{Majumder:2012sh},  this operator product, $\hat{q}$, defined over values of $k^{+} = k_{\perp}^{2}/(2 q^{-})$ 
may be expressed as the imaginary part of the more generalized operator 
product, 
\bea
\hat{Q} &=& \frac{4 \pi^{2} \alpha_{s}}{N_{c}}\!\!\! \int \frac{d^{4}y d^{4} k}{(2\pi)^{4}} 
e^{i k \cdot y} \frac{2 (q^{-})^{2}}{ \sqrt{2} q^{-} } 
 \frac{ \langle M | F^{+ \perp}(0) 
F_{\perp,}^{+}(y) | M \rangle}{ (q+k)^{2} + i \epsilon } .
\eea
As defined above, $\hat{Q}$ has a branch cut in the regime where $q^{+} \sim \lambda^{2} Q$, the discontinuity across the cut yields $\hat{q}$.
One may then define the following integral in the deep Euclidean region, 
\bea
I_{1} = \oint \frac{d q^{+}}{2 \pi i} \frac{  \hat{Q}(q^{+}) }{ \left( q^{+}  + q^{-} \right) },
\eea
 where, the contour is taken around the point $q^{+} = -q^{-}$.  
We now evaluate the integral using the method of residues, at $q^{+} = -q^{-}$. Since $q^{-}$ is very large compared to $k$, one may expand out the denominator, and the $y^{-}$ dependence in the numerator to obtain, 
\bea
I_{1} &=&  \frac{ 2 \sqrt{2} \pi^{2} \alpha_{s}}{N_{c}  q^{-} } \langle M | F^{+ \mu}_{\perp}
 \sum_{n=0}^{\infty} \left( \frac{ -i\mathcal{D}^{0} }{q^{-}}  \right)^{n}  F^{+}_{\perp, \mu}  | M \rangle.  \label{I1-simple}
\eea
The above equation represents a series of local operators, containing rising powers of derivatives in the numerator along with similarly rising 
powers of the hard scale $q^{-}$ in the denominator. In any evaluation of $I_{1}$, the presence of a large energy scale $q^{-}$
will allow one to terminate the series at a reasonably small value of $n$ while encumbering a minimal amount of error.

We can now deform the contour and evaluate it over the branch cut from $q^{+} > - \lambda^{2} Q $ to $q^{+} \rightarrow \infty$.
This yields,
\bea
I_{1} &=& \frac{4 \pi^{2} \alpha_{S}}{N_{c}}  \int d q^{+} \frac{d^{4} y d^{4} k}{ (2\pi)^{4} } e^{ik \cdot y} 
\frac{\delta \left( k^{+} + q^{+} - \frac{k_{\perp}^{2}}{ 2 q^{-}} \right) }{ 2 q^{-} }
 \frac{ \langle M | F^{+ \mu} (0) F^{+}_{\mu , }(y)  | M \rangle }{ \left( q^{+} + q^{-} \right)  }  \nn \\
&=& \int_{-\lambda^{2} Q}^{\lambda^{2} Q} d q^{+}  \frac{\hat{q} (q^{+})}{ q^{+} + q^{-}} + \int_{0}^{\infty} d q^{+} V(q^{+}). \label{I1-qhat}
\eea
The second term in the equation above, refers to the contribution to the operator above from vacuum gluon radiation, 
i.e., the Bremsstrahlung radiation of gluons from an off-shell quark. As such, it contributes only in the region where 
the virtuality of the incoming quark is time-like and is independent of the temperature of the medium. Thus when $T$ is varied, the second term above is a constant, while the first depends on the temperature of the 
medium. Comparing the two equations (\ref{I1-simple}) and (\ref{I1-qhat}), subtracting the pure vacuum contributions, one may evaluate $\hat{q}$.
In the remainder, we discuss the the evaluation of the operators in Eq.~($\ref{I1-simple}$).

\section{Euclidean space and results of calculations}

All the operator products in Eq.~(\ref{I1-simple}) represent local operators which may be easily evaluated by rotation to Euclidean space. 
Denoting a generic Minkowski space operator product as $\mathcal{D}^{>} (t) = \sum_{n} \langle n | e^{-\beta H} \mathcal{O}_{1}(t) \mathcal{O}_{2} (0) | n \rangle$, we obtain the simple relation that, 
$
\mathcal{D}^{>} (t=0) = i^{N_{t}} \Delta(\tau=0).
$
Where, $N_{t}$ denotes the number of time derivatives in $D^{>} (t)$.
Using the above relation, local operator products in Minkowski space may be obtained from local operator products in 
Euclidean space. 
\begin{figure}[h!]
\resizebox{2.25in}{2.25in}{\includegraphics{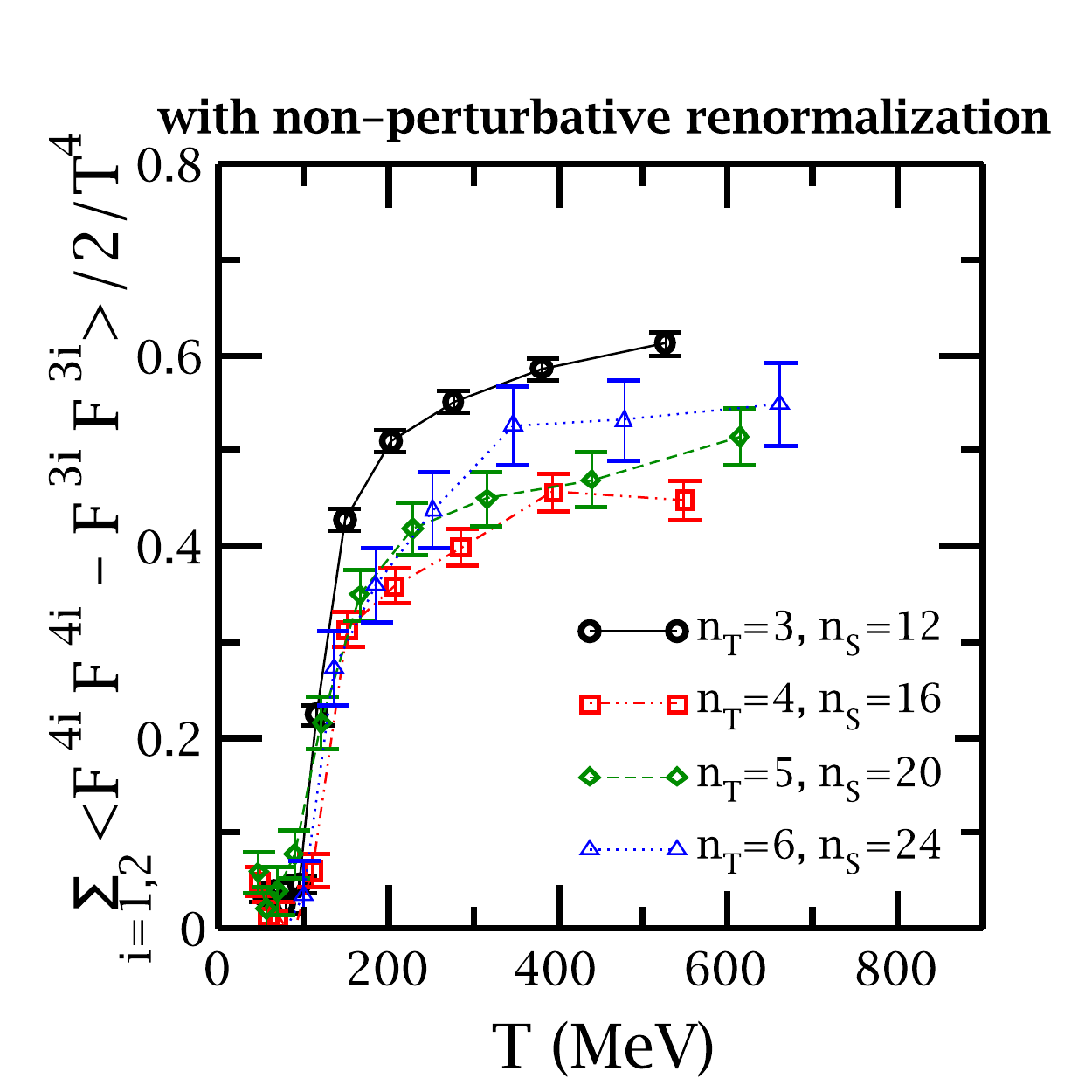}} 
\hspace{0.7in}
\resizebox{2.25in}{2.25in}{\includegraphics{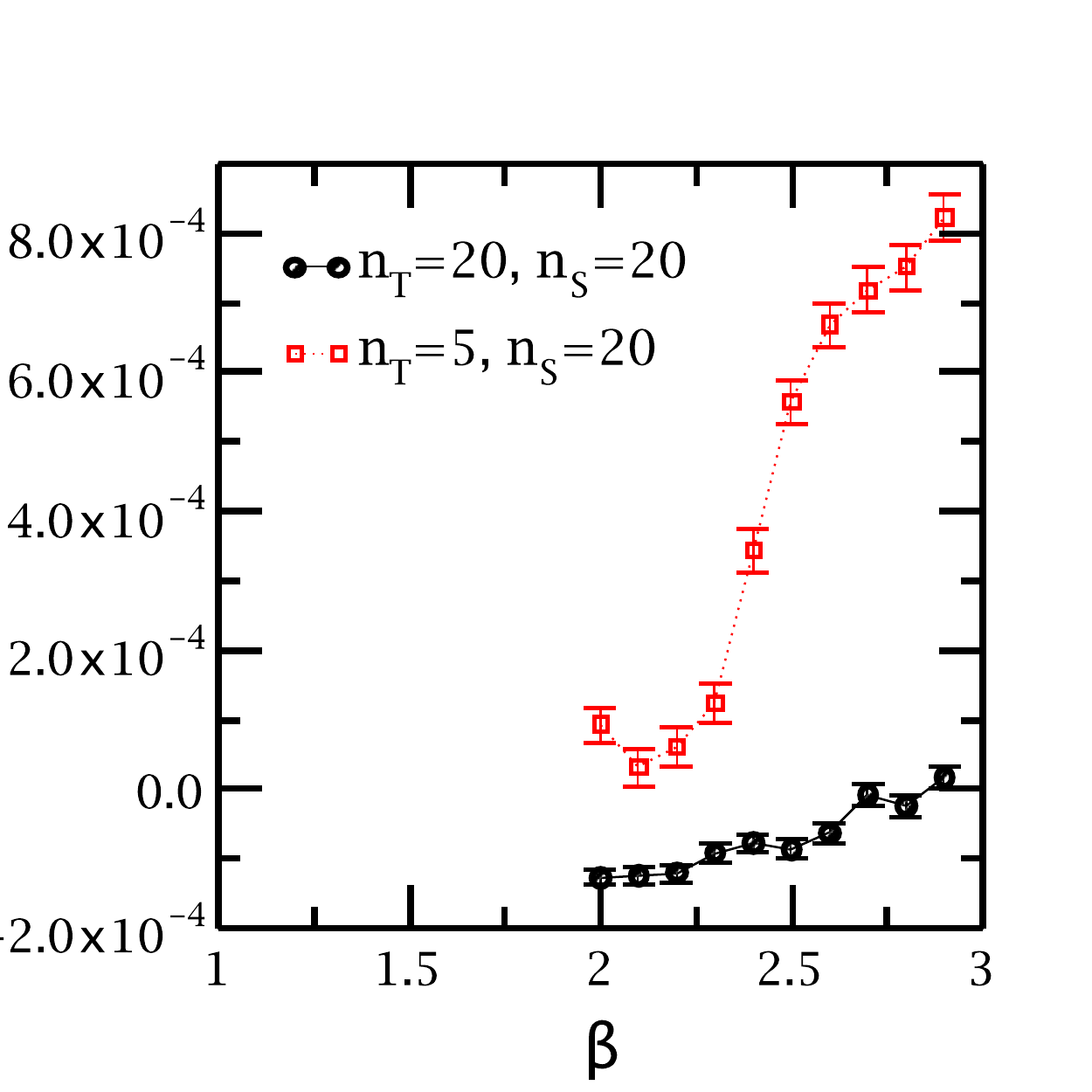}} 
    \caption{Left panel: The temperature dependence of the local operator $\langle F^{+ i} F^{+ i} \rangle$ scaled by $T^{4}$ to make it 
    dimensionless. The lattice spacing is set using a non-perturbative RG prescription. Right panel: Unscaled expectation of lattice-size-independent correlator 
$\sum_{i=1,2}a_{L}^{4}( F^{31} F^{31} - F^{41}F^{41} )/2$ at finite temperature (red squares), 
versus expectation in vacuum (black circles), as a function of $\beta = 4/g^{2}$($g$ is the bare lattice coupling). The plot is for $n_{t}=5, n_{s}=20$.}
    \label{fig1}
\end{figure}

\begin{figure}[h!]
\resizebox{2.25in}{2.25in}{\includegraphics{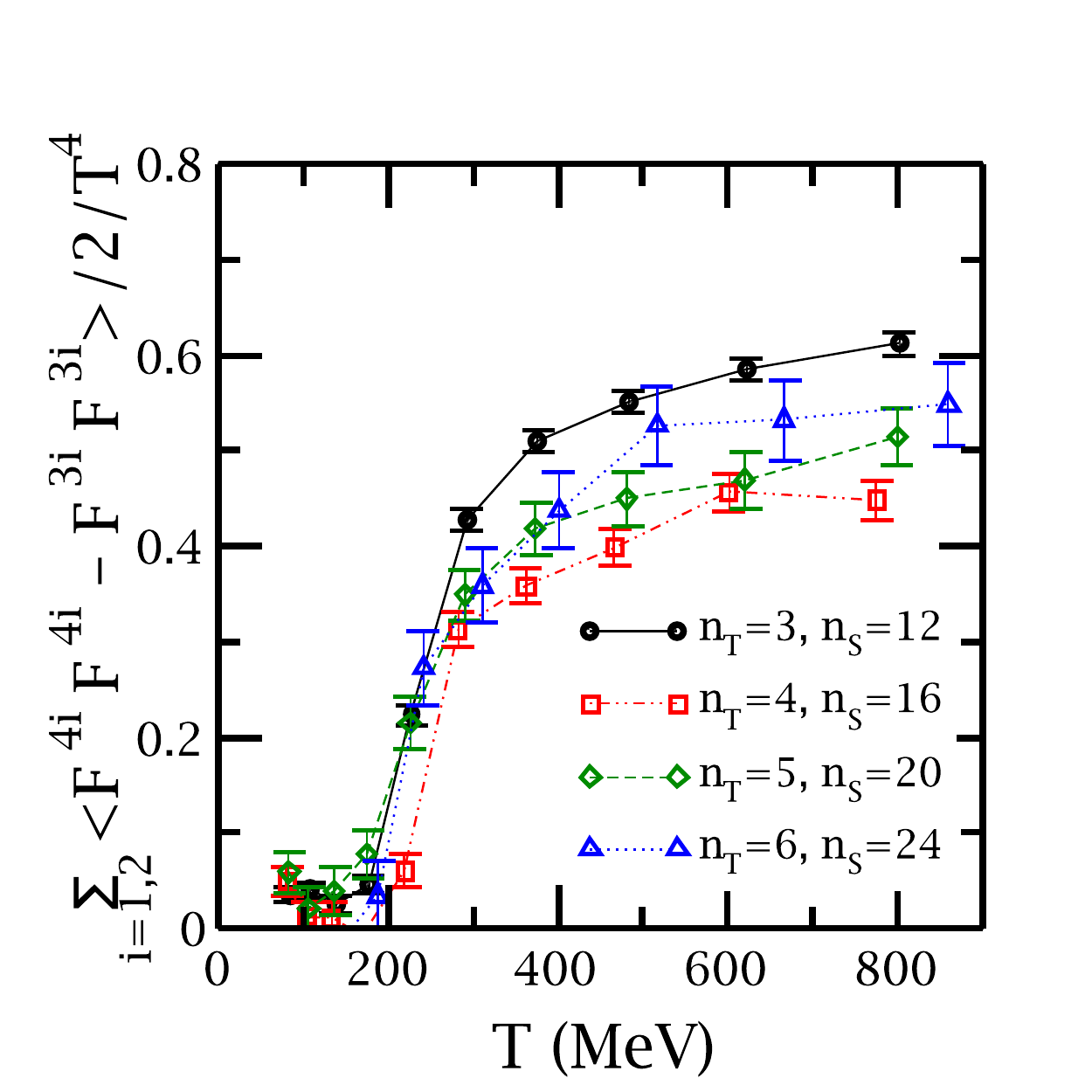}} 
\hspace{0.7in}
\resizebox{2.25in}{2.25in}{\includegraphics{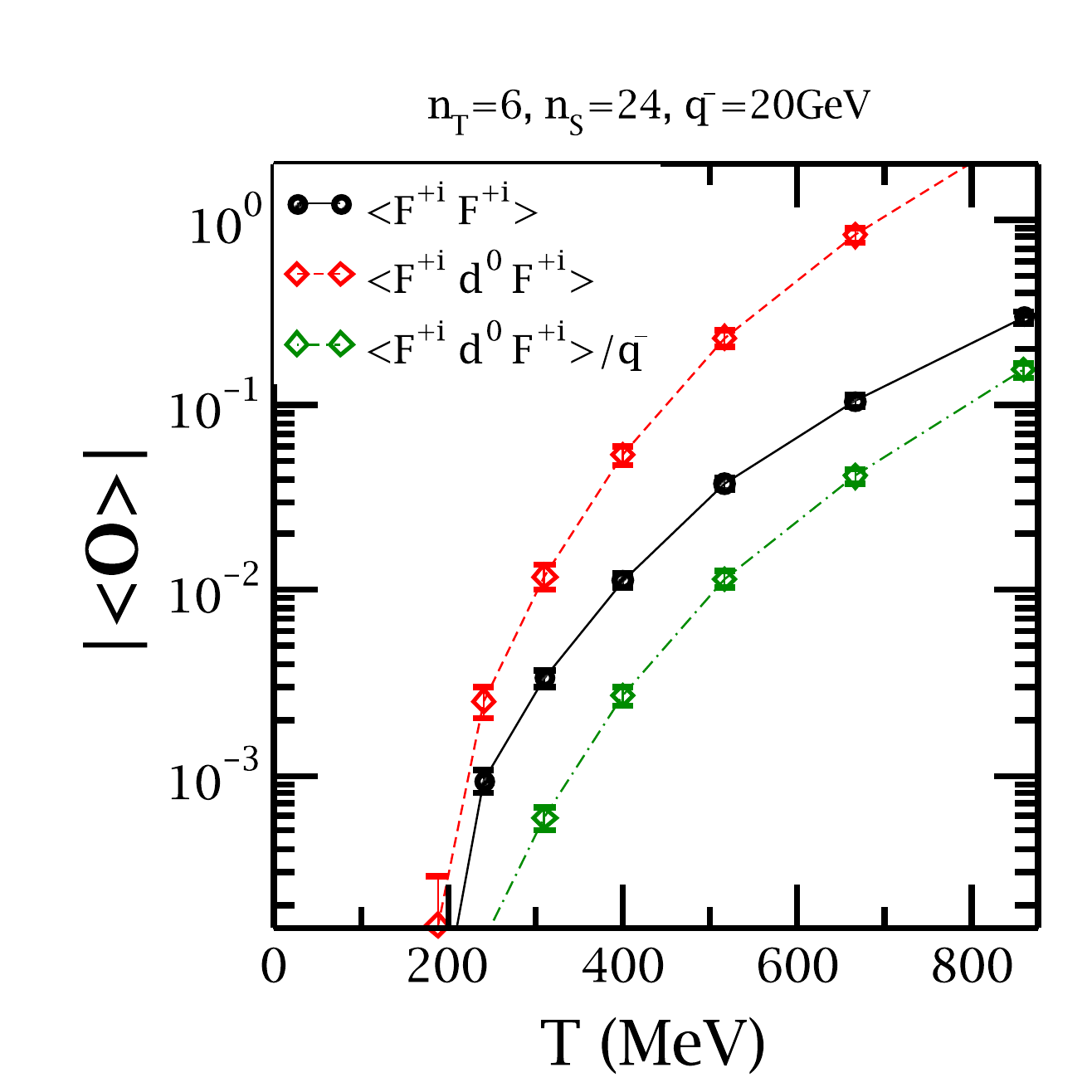}} 
    \caption{Left panel: The temperature dependence of the local operator $\langle F^{+ i} F^{+ i} \rangle$ scaled by $T^{4}$ to make it 
    dimensionless. Right panel: temperature dependence of absolute values of the local operator $\langle F^{+ i} F^{+ i} \rangle$ and the next-to-leading operator products $\langle \left[ F^{+ i} i \mathcal{D}^{4} F^{+ i} \right] \rangle$ [red diamond (dashed)] and 
    $\langle \left[ F^{+ i} i \mathcal{D}^{4} F^{+ i} \right]/q^{-} \rangle$ [green diamond (dot dashed)].}
    \label{fig2}
\end{figure}

We report results on a $(4\times n_{t})^{3}\times n_{t}$ lattice where $n_{t}$ is varied from $3$ to $6$. 
All calculations have been done with 5000 heat bath sweeps for each data point.
We have used the Wilson gauge action for SU(2)~\cite{Engels:1980ty,Creutz:1984mg}. 
The scale is set on the lattice using two different renormalization 
group (RG) formulas: The first is based on the two loop perturbative RG equation for the 
string tension~\cite{Engels:1980ty,Creutz:1984mg}, which yields the following formula for the lattice spacing,
\bea
a_{L} = \frac{1}{\Lambda_{L}} \left(\frac{11 g^{2}}{ 24 \pi^{2}}\right)^{-\frac{51}{121}} \exp \left( - \frac{12\pi^{2} }{11 g^{2}} \right), \label{lattice-spacing}
\eea
where, $g$ represents the bare lattice coupling and $\Lambda_{L}$ represents the one 
dimension-full parameter on the lattice. Comparing with the vacuum string tension, we have used $\Lambda_{L}=5.3$~MeV.   
For a lattice at finite temperature or one with $n_{t} \ll n_{s}$, the temperature is obtained as
$
T = 1 / ( n_{t} a_{L} ).
$
The results for the field-strength-field-strength correlation (FFC), $\sum_{i=1,2}( F^{3i} F^{3i} - F^{4i}F^{4i} )/2$  with this choice of 
formula for the lattice spacing are presented in the left panel of Fig~\ref{fig2}. The resulting correlation is scaled by $T^{4}$, with $T$ as obtained 
from the formula above. 

We have also set the scale using a non-perturbative approach as outlined in Ref.~\cite{Engels:1994xj}, 
where the formula for the lattice spacing is expressed as the product of that obtained from Eq.~(\ref{lattice-spacing}) and a 
non-perturbative function $\lambda(g^{2})$ which has been dialed to ensure that $T_{c}/\Lambda_{L}$ is independent of $g^{2}$. 
The results for the FFC, $\sum_{i=1,2}( F^{3i} F^{3i} - F^{4i}F^{4i} )/2$ with this next choice of 
formula for the lattice spacing are presented in the left panel of Fig~\ref{fig1}. While the scaling with lattice size is seen to be much better with 
the non-perturbative RG equation, for temperatures above $T=400$~MeV, both calculations yield the same value of the leading operator product 
in the series for $I_{1}$. In the following we will only focus on the results above a $T=400$~MeV. As a result, we will use the perturbative RG equation 
to set the scale in the evaluation of the next-to-leading terms. The field-strength-field-strength correlation, plotted in the left panels of 
Fig.~\ref{fig1} and Fig.~\ref{fig2} contain their vacuum parts, these have not been subtracted out. In the right panel of Fig.~\ref{fig1}, we plot the 
unscaled correlator as a function of the coupling on the lattice and also plot the vacuum piece. For the points above the transition, the 
vacuum contribution is seen to be a small correction, and thus was ignored in all other plots.

The next-to-leading operator product in the series for $I_{1}$ is plotted in the right panel of Fig.~\ref{fig2}.  Note, this operator has an 
extra temporal derivative and suppressed by a factor of $q^{-}$. The results of the lattice calculation of this operator are plotted, both with and 
without this large factor in the denominator ($q^{-}$ is chosen to be 20~GeV).   For temperatures below $T=600$~MeV, this next-to-leading correction 
is found to be less than 20\% of the leading term.  As a result, in the range of temperatures from $400$-$600$~MeV, the first term in the series 
expansion of $I_{1}$ can be used to obtain an estimate for $\hat{q}$. The phenomenology of these estimates is presented in Ref.~\cite{Majumder:2012sh};  
for the case of a SU(2) quark traversing a SU(2) plasma, we obtain a virtuality averaged $\bar{\hat{q}} = 0.186 \rm{GeV}^{2}/{\rm fm}$, at a $T=400$~MeV. 
Naively scaling this result with the number of colors and flavors for the 
case of an SU(3) gluon traversing an SU(3) plasma with two light flavors of quarks, we obtain $\bar{\hat{q}} = 1.3 - 3.3 $~GeV$^{2}$/fm at a $T=400$~MeV.

This work was supported by the NSF under grant no PHY-1207918, and partially by the U.S. Department of Energy under grant no. DE-SC0004286 and (within the framework of the JET 
collaboration) DE-SC0004104. Computations were carried out on the Wayne State Grid.



\bibliographystyle{00}

\end{document}